\documentclass[journal]{IEEEtran}

\usepackage{graphicx}
\usepackage{amsmath}
\usepackage{amssymb}
\usepackage[caption=false]{subfig}
\usepackage[noadjust]{cite}
\usepackage{float}
\usepackage{algorithm}
\usepackage{bibentry}
\usepackage{balance}
\usepackage{algorithm}
\usepackage{algorithmicx}
\usepackage{algpseudocode}
\usepackage{xcolor}
\usepackage{balance}
\usepackage{multirow}

\graphicspath{{img/}}

\newtheorem{theorem}{Theorem}

\DeclareMathOperator{\Tr}{\rm tr}

\begin{document}

\bstctlcite{IEEEexample:BSTcontrol}

\title{Multiuser Precoding for Sum-Rate Maximization in Relay-Aided mmWave Communications}

\author{Ahmet Zahid Yal\c{c}{\i}n and Yavuz Yap{\i}c{\i}}

\maketitle

\begin{abstract}
Relay-aided transmission is envisioned as a key strategy to combat severe path loss and link blockages emerging as unique challenges in millimeter-wave (mmWave) communications. This work considers a relay-aided multiuser mmWave communications scenario aiming at maximizing the sum rate through optimal transmit and relay precoder design. We propose a novel joint precoder design strategy, which exploits weighted minimum mean-square error (WMMSE) using its equivalency to sum-rate maximization. We obtain closed form expressions of transmit and relay precoders, and propose to compute them through alternating-optimization iterations without having to resort to complicated numerical optimization techniques. Numerical results verify the superiority of the proposed precoding strategy as compared to conventional precoding schemes.
\end{abstract}

\begin{IEEEkeywords}
Amplify-and-forward (AF) relaying, convex optimization, sum rate, minimum mean-square error (MMSE).
\end{IEEEkeywords}

\section{Introduction} \label{sec:intro}

The millimeter-wave (mmWave) communication has recently received enormous attention as a promising solution to the spectrum scarcity problem in conventional sub-6 GHz frequency band \cite{Roh2014mmW}. This vastly unused spectrum, however, comes with a set of unique challenges such as severe propagation loss and link blockages. As an attempt to combat these challenges, relay-aided communications schemes have recently been revisited for mmWave channels. Along with the use of multiple antennas, the existing literature on relay-aided networks is focused more on the \textit{uncorrelated} multiple-input multiple-output (MIMO) channels {\color{black}\cite{7974807,7817896,8013765,7869326,7581035,Kong2017_JointMMSE, Joung2017_LinearPrecoder, Cumanan2017_RobustMSE, Shim2018_BeamformingDesign,8357941}}, with relatively limited attention to \textit{correlated} mmWave channels having mostly \textit{point-to-point} transmission structure \cite{Wang2018PreDes, Dai2018RelHyb, Meng2019OnPre}.

In particular, {\color{black} \cite{7974807} considers precoding design problem for a full-duplex (FD) amplify-and-forward (AF) relaying scheme which aims at maximizing sum rates without any source node processing. A joint source and relay precoding strategy is studied in \cite{7817896} to mitigate the loop interference of an FD scheme in an AF relaying network. As a follow up work, the authors investigate nonlinear transceiver design in \cite{8013765} for the same setting, where the source and relay precoders are designed assuming (nonlinear) successive-interference-cancellation (SIC) receiver. \cite{7869326} investigates the simultaneous wireless information and power transfer (SWIPT) in a MIMO AF relaying system, and proposes a source and relay precoding scheme assuming an energy-harvesting relay. The optimal transmit and relay precoders (together with receiver equalizer) are investigated in \cite{7581035} for an AF relaying scheme which aims at minimizing the probability of error. \cite{8357941} considers the source and relay precoder design problem for FD AF relaying systems, where SWIPT is enabled at the destination.}

{\color{black}The precoder design problem for transmit and relay nodes is studied in \cite{Kong2017_JointMMSE} with a non-negligible direct link between the transmitter and users.} A two-way relay structure is considered in \cite{Joung2017_LinearPrecoder} for which the relay precoder is designed aiming at maximizing the spectral efficiency. An iterative minimum mean-square error (MMSE) precoder is proposed in \cite{Cumanan2017_RobustMSE} which is argued to be robust against channel state information (CSI) uncertainties. A multi-hop MIMO transmission is suggested in \cite{Shim2018_BeamformingDesign} where a mean-square error (MSE) beamforming design is studied. In \cite{Wang2018PreDes}, precoder design is studied for a two-hop relay network so as to maximize average capacity. The work in \cite{Dai2018RelHyb} proposes an MMSE-based iterative successive approximation (ISA) algorithm to design a hybrid beamformer. An FD relaying scheme is studied in \cite{Meng2019OnPre}, and an self-interference cancellation precoding is proposed accordingly.

In this work, we consider an AF based relay-aided MIMO transmission scenario for \textit{multiuser} mmWave communications. In particular, we propose a novel framework for joint transmit and relay precoder design problem which aims at maximizing the sum rate. In order to solve the respective optimization, we resort to an equivalent optimization problem involving weighted MMSE (WMMSE) of message decoding, and derive the closed-form expressions for the transmit and relay precoders together with the Lagrange multipliers. We further propose an alternating-optimization algorithm to obtain the precoders through iterations without having to carry out numerically expensive optimization techniques. The sum-rate performance of the proposed scheme is verified by the extensive numerical results to achieve superior performance as compared to conventional precoding schemes.

{\color{black}The rest of the paper is as follows. Section \ref{sec:system} describes the system model under consideration. The achievable rates are presented in Section \ref{sec:rate_receiver} together with the derivation of equivalent MMSE receiver. Section \ref{sec:optimal_precoder} discusses the optimal sum-rate maximizing precoders, and Section \ref{sec:proposed_precoder} proposes a novel iterative algorithm to compute the optimal transmit and relay precoders. The numerical results are presented in Section \ref{sec:results}, and the paper concludes with some remarks in Section \ref{sec:conclusion}. 

\textit{Notation:} $(\cdot)^{\rm H}$, ${\rm tr}(\cdot)$, and $|\cdot|$ are Hermitian, trace, and absolute value operators, respectively. $[a,b]$ is a continuous interval between (inclusive) $a$ and $b$. $\mathbb{C}^{K{\times}L}$ denotes the set of $K{\times}L$ complex numbers. $\textbf{I}_N$ is the identity matrix of size $N{\times}N$. $\mathbb{E}\{\cdot\}$ denotes statistical expectation.}

\section{System Model} 
\label{sec:system}

We consider a relay-aided mmWave communications scenario in which a base station (BS) equipped with $N_\mathsf{t}$ antennas communicates with $K$ single-antenna users through a relay node having $N_\mathsf{r}$ antennas. The users are represented by an index set $\mathcal{K} \,{=}\, \{1,\dots,K\}$ such that $K \,{\leq}\, \mathsf{min}\{N_\mathsf{t},N_\mathsf{r}\}$. We assume that users are uniformly distributed over the horizontal plane each with a line-of-sight (LoS) distance (from the BS) in $[d_\mathsf{min},d_\mathsf{max}]$, and an angular position in $[\theta_\mathsf{min},\theta_\mathsf{max}]$. We also assume that the relay is off the BS by $d_\mathsf{r}$, and is aligned with the boresight of the BS propagation pattern.

\subsection{\color{black}Channel Model} \label{subsec:channel_model}

\color{black}
The mmWave channel $\textbf{H}_\mathsf{sr} \,{\in}\, \mathbb{C}^{N_\mathsf{r}{\times}N_\mathsf{t}}$ between the BS (source) and relay is given as follows
\begin{align} \label{eq:channel_SR}
\textbf{H}_\mathsf{sr} = 
\sqrt{ \frac{ N_\mathsf{t} N_\mathsf{r} }{ M_\mathsf{sr} L_\mathsf{sr}} } \sum \limits_{m=1}^{M_\mathsf{sr}} \sum \limits_{\ell=1}^{L_\mathsf{sr}} \frac{ \alpha_{m,\ell}^\mathsf{sr} }{ \sqrt{ \mathsf{PL} \left(d_\mathsf{sr}\right) } } \, \textbf{a}(N_\mathsf{r}, \theta_{m,\ell}^\mathsf{sr} ) \textbf{a}(N_\mathsf{t}, \phi_{m,\ell}^\mathsf{sr} )^{\rm H}, 
\end{align} 
where $M_\mathsf{sr}$ and $L_\mathsf{sr}$ are the number of clusters and multipath components, respectively,  $\mathsf{PL}(d_\mathsf{sr})$ is the path loss over $d_\mathsf{sr}$ being the line-of-sight (LoS) distance between the BS and relay, and $\alpha_{m,\ell}^\mathsf{sr}$ is the gain of the $\ell$-th multipath in the $m$-th cluster being standard complex Gaussian. In addition, $\theta_{m,\ell}^\mathsf{sr}$ and $\phi_{m,\ell}^\mathsf{sr}$ are the uniformly-distributed angle-of-arrival (AoA) and angle-of-departure (AoD) of the $\ell$-th multipath in the $m$-th cluster, respectively, and $\textbf{a}(N,\varphi)$ stands for the array steering vector of size $N{\times}1$ with the $n$-th element given for the uniform linear array (ULA) assumption as follows
\begin{align} \label{eq:steer}
\Big[ \textbf{a}(N,\varphi) \Big]_n =  \frac{1}{\sqrt{M}} {\rm exp} \left\lbrace {-}j2\pi \frac{d}{\lambda} \left( n{-}1\right) \cos\varphi \right\rbrace,
\end{align}
for $n \,{=}\, 1,\dots,N$, where $\varphi$ is the angle of interest (i.e., AoA or AoD), $d$ is the antenna element spacing, and $\lambda$ is the wavelength of the carrier frequency.

The mmWave channel ${\textbf{h}_k} \,{\in}\, \mathbb{C}^{1{\times}N_\mathsf{r}}$ between the relay and $k$-th user (destination) is similarly defined as follows
\begin{align} \label{eq:channel_SD_RD}
\textbf{h}_k = \sqrt{ \frac{ N_\mathsf{r} }{ M_\mathsf{rd} L_\mathsf{rd} } } \sum \limits_{m=1}^{M_\mathsf{rd}} \sum \limits_{\ell=1}^{L_\mathsf{rd}} \frac{ \alpha_{m,\ell}^k }{ \sqrt{ \mathsf{PL} \left(d_k\right) } } \, \textbf{a}( N_\mathsf{r}, \phi_{m,\ell}^k )^{\rm H},
\end{align} 
where $d_k$ is the LoS distance between the relay and $k$-th user, $\alpha_{m,\ell}^k$ is the gain of the $\ell$-th multipath in the $m$-th cluster being standard complex Gaussian, and $\phi_{m,\ell}^k$ is the respective AoD. 

\color{black}

\subsection{\color{black}Transmission Scheme}

The BS intends to send unicast message symbol $s_k$ to the $k$-th user in two phases following AF relaying scheme. In particular, the BS transmits only to the relay node in the first phase (assuming the users are blocked due to the mmWave propagation characteristics), and the relay node retransmits the received signal to the users in the second phase (after linear processing).   

In the first phase, the overall unicast message vector $\textbf{s} \,{=}\, [s_1 \dots s_K]^{\rm T}$ ${\in}\, \mathbb{C}^{K{\times}1}$ with $\mathbb{E}\{\textbf{s}\textbf{s}^{\rm H}\} \,{=}\, \textbf{I}_K$ is transmitted by the BS using a linear precoder $\textbf{F} \,{=}\, [\textbf{f}_1 \dots \textbf{f}_K]$ ${\in}\, \mathbb{C}^{N_\mathsf{t}{\times}K}$, where $\textbf{f}_k$ ${\in}\, \mathbb{C}^{N_\mathsf{t}{\times}1}$ is the precoder vector for the $k$-th unicast message. The received signal at the relay node is given as
\begin{align} \label{eq:observation_r}
 \textbf{y}_\mathsf{r} &= \sqrt{\rho_\mathsf{s}} \textbf{H}_\mathsf{sr} \textbf{F} \textbf{s} + \textbf{n} = \sqrt{\rho_\mathsf{s}} \textbf{H}_\mathsf{sr} \sum_{k \in \mathcal{K}} \textbf{f}_k s_k + \textbf{n} ,
\end{align}
where $\rho_\mathsf{s}$ is the control parameter for the transmit power at the BS, and $\textbf{n}$ is the complex Gaussian noise with zero mean and covariance $\sigma^2_\mathsf{r} \textbf{I}_{N_\mathsf{r}}$. Assuming that long-term power budget at the BS is $\mathsf{P}_\mathsf{bs}$, the respective power constraint for $\textbf{F}$ is 
\begin{align} \label{eq:power_const_bs}
\Tr \left( \textbf{F} \textbf{F}^{\rm H} \right) = \sum_{k \in \mathcal{K}} \Tr \left( \textbf{f}_k \textbf{f}_k^{\rm H} \right) \leq  \mathsf{P}_\mathsf{bs} / \rho_\mathsf{s}.
\end{align}

In the second phase, the relay node forwards the received signal $\textbf{y}_\mathsf{r}$ to the users by the precoder $\textbf{G} \,{\in}\, \mathbb{C}^{N_\mathsf{r}{\times} N_\mathsf{r}}$, which acts to amplify the incoming signal in power, and to suppress multiuser interference (aiming at maximizing the sum rate). The received signal at the $k$-th user is given by
\begin{align}
 y_k &= \sqrt{ \rho_\mathsf{r} } \textbf{h}_k \textbf{G} \textbf{y}_\mathsf{r} + w_k , \\
 &= \sqrt{ \rho_\mathsf{s} \rho_\mathsf{r} } \textbf{h}_k \textbf{G} \textbf{H}_\mathsf{sr}  \sum_{k \in \mathcal{K}} \textbf{f}_k s_k + \sqrt{ \rho_\mathsf{r} } \textbf{h}_k \textbf{G} \textbf{n} + w_k, \label{eq:observation_u}
\end{align} 
where $\rho_\mathsf{r}$ is the control parameter for the transmit power at the relay node, and $w_k$ is the complex Gaussian noise with zero mean and variance $\sigma^2_\mathsf{d}$. The long-term power budget at the relay node is $\mathsf{P}_\mathsf{re}$, and the power constraint for the relay precoder is  
\begin{align} \label{eq:power_const_re}
\rho_\mathsf{s} \sum_{k \in \mathcal{K}} \Tr\left( \textbf{G} \textbf{H}_\mathsf{sr}  \textbf{f}_k\textbf{f}_k^{\rm H} \textbf{H}_\mathsf{sr} ^{\rm H}\textbf{G}^{\rm H} \right) + \sigma^2_\mathsf{r} \Tr\left( \textbf{G}\textbf{G}^{\rm H} \right) \leq \mathsf{P}_\mathsf{re} / \rho_\mathsf{r}.
\end{align}

\section{Achievable Rates and MMSE Receiver} \label{sec:rate_receiver}

The one-to-one correspondence between mutual information and MMSE is established in \cite{Cioffi2008_WSR}. In this section, we consider the achievable rates and the MMSE receiver to establish such a relation. To this end, we first consider the receiver operation in which each user decodes its own unicast message. The respective rate of the $k$-th user message can be given as follows
\begin{align} \label{eq:rate_user}
\mathsf{R}_k &= \log \left( 1 + \frac{\rho_\mathsf{s}\rho_\mathsf{r}}{\sigma^{2}_k} \, \textbf{h}_k \textbf{G} \textbf{H}_\mathsf{sr}  \textbf{f}_k \textbf{f}_k^{\rm H} \textbf{H}_\mathsf{sr} ^{\rm H} \textbf{G}^{\rm H}  \textbf{h}_k^{\rm H} \right),
\end{align}
where $\sigma^2_k$ is the \textit{effective} noise variance given as
\begin{align}\label{eq:effective_noise_var}
\!\!\!\sigma^2_k \,{=} \sum_{\substack{i\ne k}} \!  \rho_\mathsf{s} \rho_\mathsf{r} \textbf{h}_k \textbf{G} \textbf{H}_\mathsf{sr} \textbf{f}_i \textbf{f}_i^{\rm H} \textbf{H}_\mathsf{sr} ^{\rm H} \textbf{G}^{\rm H}  \textbf{h}_k^{\rm H} {+} \rho_\mathsf{r} \sigma^2_\mathsf{r} \textbf{h}_k \textbf{G} \textbf{G}^{\rm H}  \textbf{h}_k^{\rm H} {+} \sigma^2_\mathsf{d},
\end{align}
which specifically depends on the user index $k \,{\in}\, \mathcal{K}$.

At the receive side, the $k$-th user employs a receiver $V_k$ to process its received signal $y_k$ aim at obtaining an estimate of its unicast message $s_k$ as $\hat{s}_k \,{=}\, V_k y_k$. The respective MSE is
\begin{align}
\varepsilon_k \! &= \mathbb{E} \left\lbrace | s_k - V_k y_k|^2 \right\rbrace \label{eq:mse_definition} \\
& \hspace{-0.0in} = \! \bigg( \! \sum_{i \in \mathcal{K}} \rho_\mathsf{s} \rho_\mathsf{r} \textbf{h}_k \textbf{G} \textbf{H}_\mathsf{sr}  \textbf{f}_i \textbf{f}_i^{\rm H} \textbf{H}_\mathsf{sr} ^{\rm H} \textbf{G}^{\rm H}  \textbf{h}_k^{\rm H} {+} \sigma^2_\mathsf{r} \rho_\mathsf{r} \textbf{h}_k \textbf{G} \textbf{G}^{\rm H}  \textbf{h}_k^{\rm H} \bigg) V_k^* V_k \nonumber \\
& \hspace{-0.2in} + \sigma^2_\mathsf{d} V_k^* V_k {-} \sqrt{\rho_\mathsf{s} \rho_\mathsf{r}} \left( V_k \textbf{h}_k \textbf{G} \textbf{H}_\mathsf{sr}  \textbf{f}_k + V_k^* \textbf{f}_k^{\rm H} \textbf{H}_\mathsf{sr}^{\rm H} \textbf{G}^{\rm H} \textbf{h}_k^{\rm H} \right) {+} 1 . \label{eq:mse_user_k}
\end{align}
The optimal receiver for the $k$-th user minimizing the MSE,
referred to as MMSE receiver, is obtained by taking derivative of \eqref{eq:mse_user_k} with respect to $V_k$ and finding the respective root, which is given after straightforward manipulations as follows
\begin{align}
V_k^\mathsf{MMSE} &= \frac{\sqrt{\rho_\mathsf{s}\rho_\mathsf{r}} \, \textbf{f}_k^{\rm H} \textbf{H}_\mathsf{sr}^{\rm H} \textbf{G}^{\rm H}  \textbf{h}_k^{\rm H}} {\rho_\mathsf{s}\rho_\mathsf{r} \, \textbf{h}_k \textbf{G} \textbf{H}_\mathsf{sr}  \textbf{f}_k \textbf{f}_k^{\rm H} \textbf{H}_\mathsf{sr} ^{\rm H} \textbf{G}^{\rm H}  \textbf{h}_k^{\rm H} + \sigma^2_k }, \label{eq:mmse_receiver}
\end{align}
where $\sigma^2_k$ is the effective noise for the $k$-th user given in \eqref{eq:effective_noise_var}. 

When the MMSE receiver of \eqref{eq:mmse_receiver} is employed at the $k$-th user, the resulting MSE of \eqref{eq:mse_user_k} becomes
\begin{align}
\varepsilon_k^\mathsf{min} &=\left( 1 + \frac{\rho_\mathsf{s}\rho_\mathsf{r}}{\sigma^2_k} \textbf{f}_k^{\rm H} \textbf{H}_\mathsf{sr} ^{\rm H} \textbf{G}^{\rm H}  \textbf{h}_k^{\rm H} \textbf{h}_k \textbf{G} \textbf{H}_\mathsf{sr}  \textbf{f}_k \right)^{-1}. \label{eq:mmse_value}
\end{align}
We furthermore realize that the achievable rate in \eqref{eq:rate_user} is related to the MMSE value in \eqref{eq:mmse_value} through the following expression
\begin{align}
\mathsf{R}_k &= - \log \left( \varepsilon_k^\mathsf{min} \right). \label{eq:rate_mmse_relation}
\end{align}

\section{Optimal Precoders for Sum Rate Maximization} \label{sec:optimal_precoder}

Our ultimate goal is to derive the optimal transmit and relay precoders $\textbf{F}$ and $\textbf{G}$ such that the sum rate is maximized subject to the power constrains at the BS and relay node. The respective optimization problem is defined as
\begin{IEEEeqnarray}{rl}
\max_{\textbf{F}, \textbf{G}}
&\qquad \sum_{k \in \mathcal{K}} \mathsf{R}_k, \label{eq:optimization_WSR} \\
\text{s.t.}
&\qquad \eqref{eq:power_const_bs},  \eqref{eq:power_const_re}.
\IEEEyessubnumber \label{eq:optimization_WSR_power}
\end{IEEEeqnarray}

The optimization problem in \eqref{eq:optimization_WSR} is seemingly non-convex due to the non-convex rate expression of \eqref{eq:rate_user}. We therefore resort to an equivalent optimization problem exploiting the relation between the rates and MMSE given by \eqref{eq:rate_mmse_relation}. To this end, we define an \textit{augmented} weighted MSE (WMSE) expression for the $k$-th user as follows 
\begin{align}
\xi_k &= v_k \, \varepsilon_k \,{-}\, \log \! \left( v_k \right) ,\label{eq:wmse_augmented}
\end{align}
where $v_k$ is the nonzero weight coefficient of the MSE expression for the $k$-th user. 

Note that the augmented WMSE $\xi_k$ in \eqref{eq:wmse_augmented} is convex in the MSE receiver $V_k$ through the MSE expression $\varepsilon_k$ defined in \eqref{eq:mse_definition}. In order to find the optimal weight coefficient $v_k^\mathsf{opt}$ and the receiver $V_k^\mathsf{opt}$ that minimize the augmented WMSE, we compute partial derivatives of \eqref{eq:wmse_augmented} which yields
\begin{align}
    \frac{\partial \xi_k}{\partial V_k} &= v_k \frac{\partial \varepsilon_k}{\partial V_k} \rightarrow V_k^\mathsf{opt} = V_k^\mathsf{MMSE} , \label{eq:wmse_optimal_receiver}\\
    \frac{\partial \xi_k}{\partial v_k} &= \varepsilon_k - \frac{1}{v_k} \rightarrow v_k^\mathsf{opt} = \frac{1}{ \varepsilon_k^\mathsf{min} } , \label{eq:wmse_optimal_weight}
\end{align}
where \eqref{eq:wmse_optimal_receiver} makes use of the fact that the root of $\partial \varepsilon_k / \partial V_k$ is the MMSE receiver in \eqref{eq:mmse_receiver}, and \eqref{eq:wmse_optimal_weight} exploits the finding from \eqref{eq:wmse_optimal_receiver} that the optimal value of $\varepsilon_k$ is given by \eqref{eq:mmse_value} due to the optimal receiver still being MMSE for this particular problem. 

The minimum of the augmented WMSE, referred to as WMMSE, is obtained by the optimal receiver of \eqref{eq:wmse_optimal_receiver} and the optimal weight of \eqref{eq:wmse_optimal_weight} as follows
\begin{align}
\xi_k^\mathsf{min} &= 1 - \mathsf{R}_k , \label{eq:wmse_augmented_min}
\end{align}
which makes use of the relation between the MMSE and achievable rate given by \eqref{eq:rate_mmse_relation}. We therefore formulate the equivalent optimization problem which minimizes the WMSE (instead of maximizing sum rate as in \eqref{eq:optimization_WSR}) as follows
\begin{IEEEeqnarray}{rl}
\min_{\textbf{F}, \textbf{G}}
&\qquad \sum_{k \in \mathcal{K}} \xi_k^\mathsf{min}  \label{eq:optimization_WMMSE} \\
\text{s.t.}
&\qquad \eqref{eq:power_const_bs}, \eqref{eq:power_const_re}.
\IEEEyessubnumber \label{eq:optimization_WMMSE_power}
\end{IEEEeqnarray}
In the following, we describe an alternating-optimization approach to find the optimal transmit and relay precoders maximizing the sum rate. 

\section{Iterative Precoder Design} 
\label{sec:proposed_precoder}

The optimization problem in \eqref{eq:optimization_WSR} that aims at maximizing the sum rate is non-convex, and requires numerical optimization techniques to solve. The equivalent WMMSE problem of \eqref{eq:optimization_WMMSE}, however, leads to a low-complexity intuitive algorithm which makes use of the convexity of the objective function in receiver weights. The WMMSE problem is composed of three parts: the transmit precoders, relay precoder, and MSE receivers. One intuitive way to solve \eqref{eq:optimization_WMMSE} is therefore through an alternating-optimization scheme (i.e., keeping two of these constituent parts the same while optimizing the other one).

Towards this end, we first define a new optimization problem based on \eqref{eq:optimization_WMMSE}, which employs the general augmented WMSE $\xi_k$ of \eqref{eq:wmse_augmented} as the objective function (instead of the WMMSE $\xi_k^\mathsf{min}$ as \eqref{eq:optimization_WMMSE} uses), or equivalently considering the MSE $\varepsilon_k$ of  \ref{eq:mse_user_k} (instead of using the MMSE $\varepsilon_k^{min}$ of \eqref{eq:mmse_value}). This new optimization therefore assumes general receivers (along with the general weight coefficients $v_k$) instead of specifically assuming the MMSE receivers of \eqref{eq:mmse_receiver} (with the WMMSE-optimal weight coefficients $v_k^\mathsf{opt}$), and is given as
\begin{IEEEeqnarray}{rl}
\min_{\substack{\textbf{F}, \textbf{G}, V_k, v_k}}
&\qquad \sum_{k \in \mathcal{K}} \xi_k  \label{eq:optimization_WMMSE_2} \\
\text{s.t.}
&\qquad \eqref{eq:power_const_bs}, \eqref{eq:power_const_re}.
\IEEEyessubnumber \label{eq:optimization_WMMSE_power_2}
\end{IEEEeqnarray}

Thanks to the objective function being convex in the precoders and general receivers, this optimization problem can be solved considering the Lagrangian function given as
\begin{align}
\mathcal{L} \left( \textbf{F}, \textbf{G}, V_k \right) &= \sum_{k \in \mathcal{K}} \xi_k + \beta_1 J_1 + \beta_2 J_2 , \label{eq:lagrangian}
\end{align}
where $\beta_1$, $\beta_2$, and $\eta_k$ are the Lagrange multipliers, and $J_1$ and $J_2$ are the penalty for the power constraint at the BS and the relay node, respectively, which are given as 
\begin{align}
    \hspace{-0.1in} J_1 &= \sum_{k \in \mathcal{K}} \Tr (\textbf{f}_k\textbf{f}_k^{\rm H}) - \frac{ \mathsf{P}_\mathsf{bs} }{ \rho_\mathsf{s} } , \\
    \hspace{-0.1in} J_2 &= \rho_\mathsf{s} \sum_{k \in \mathcal{K}} \Tr(\textbf{G} \textbf{H}_\mathsf{sr}  \textbf{f}_k\textbf{f}_k^{\rm H} \textbf{H}_\mathsf{sr}^{\rm H}\textbf{G}^{\rm H} ) + \sigma^2_\mathsf{r} \Tr( \textbf{G}\textbf{G}^{\rm H} ) - \frac{\mathsf{P}_\mathsf{re} }{ \rho_\mathsf{r} } .
\end{align}
Considering the Karush–Kuhn–Tucker (KKT) conditions, the desired precoders are given in the following theorem.

\begin{theorem}\label{theorem}
Assuming the relay-aided communications scenario with the observation models at the BS and relay node given by \eqref{eq:observation_r} and \eqref{eq:observation_u}, respectively, along with the power constraints \eqref{eq:power_const_bs} and \eqref{eq:power_const_re}, the optimal precoders are given as
\begin{align}
\textbf{f}_k &= \sqrt{ \rho_\mathsf{s} \rho_\mathsf{r} } v_k V_k^* \bigg[\beta_1 \textbf{I}_{N_\mathsf{t}} + \beta_2 \rho_\mathsf{s} \textbf{H}_\mathsf{sr} ^{\rm H} \textbf{G}^{\rm H} \textbf{G} \textbf{H}_\mathsf{sr}   \nonumber \\
& \quad + \rho_\mathsf{s} \rho_\mathsf{r} \sum_{i \in \mathcal{K}}  v_i |V_i|^2  \textbf{H}_\mathsf{sr} ^{\rm H} \textbf{G}^{\rm H}  \textbf{h}_i^{\rm H} \textbf{h}_i  \textbf{G} \textbf{H}_\mathsf{sr}  \bigg]^{-1}  \textbf{H}_\mathsf{sr} ^{\rm H} \textbf{G}^{\rm H}  \textbf{h}_k^{\rm H}, \label{eq:precoder_optimal_bs}\\
\textbf{G} &= \sqrt{ \rho_\mathsf{s} \rho_\mathsf{r} } \left( \rho_\mathsf{r} \sum_{i \in \mathcal{K}} v_i |V_i|^2 \textbf{h}_i^{\rm H}  \textbf{h}_i + \beta_2 \textbf{I}_K \right)^{-1} \nonumber \\
& \quad \times \left( \sum_{i \in \mathcal{K}} \! v_i V_i^* \textbf{h}_i^{\rm H}   \textbf{f}_i^{\rm H}\textbf{H}_\mathsf{sr} ^{\rm H} \right) \!\! \left( \rho_\mathsf{s} \textbf{H}_\mathsf{sr}  \textbf{F}  \textbf{F}^{\rm H} \textbf{H}_\mathsf{sr} ^{\rm H}  \,{+}\, \sigma^2_\mathsf{r}  \textbf{I}_{N_\mathsf{r}} \right)^{-1} \!\!\! . \label{eq:precoder_optimal_re}
\end{align}
for $k \,{=}\, 1,\dots,K$, where $V_k$ is obtained by \eqref{eq:mmse_receiver}, and
\begin{align}
\beta_2 &= \frac{ \rho_\mathsf{r} \sigma^2_\mathsf{d} }{ \mathsf{P}_\mathsf{re} } \sum_{i \in \mathcal{K}} v_i |V_i|^2, \label{eq:beta2}\\
\beta_1 &= \frac{\rho_\mathsf{s} \sigma^2_\mathsf{r} }{ \mathsf{P}_\mathsf{bs} } \bigg[  \rho_\mathsf{r} \sum_{i \in \mathcal{K}} v_i |V_i|^2 \Tr \left( \textbf{h}_i^{\rm H} \textbf{h}_i \textbf{G} \textbf{G}^{\rm H}  \right) + \beta_2 \Tr \left( \textbf{G} \textbf{G}^{\rm H} \right) \bigg]. \label{eq:beta1}
\end{align}
\end{theorem}
\begin{IEEEproof}
See Appendix~\ref{appendix}.
\end{IEEEproof}

\begin{algorithm}[!t]
    \caption{Proposed WMMSE-Based Algorithm}
    \label{algorithm}
    \begin{algorithmic}[1]
        \State \textbf{Initialize:} $\epsilon$, {\color{black}$n_\mathsf{max}$,} $\textbf{F}$, $\textbf{G}$,
        $n \gets 1$, $\xi_\mathsf{t}^{(0)} \gets \infty$, $\xi_\mathsf{t}^{({-}1)} \gets 0$
        \While{$\big| \xi_\mathsf{t}^{(n{-}1)} - \xi_\mathsf{t}^{(n{-}2)}
        \big| > \epsilon$ {\color{black} \textbf{or} $n > n_\mathsf{max}$} } 
            \State Compute $V_k$ by \eqref{eq:mmse_receiver} for given $\textbf{F}$ and $\textbf{G}$
            \State Compute $\varepsilon_k$ by \eqref{eq:mse_user_k} for given $\textbf{F}$ and $\textbf{G}$
            \State Compute $v_k^\mathsf{opt}$ by \eqref{eq:wmse_optimal_weight}
            \State Compute $\beta_2$ by \eqref{eq:beta2}
            \State Update $\textbf{G}$ by \eqref{eq:precoder_optimal_re} for given $\textbf{F}$ and $V_k$
            \State Compute $\beta_1$ by \eqref{eq:beta1}
            \State Update $\textbf{F}$ by \eqref{eq:precoder_optimal_bs} for given $\textbf{G}$ and $V_k$
            \State $\xi_\mathsf{t}^{(n)} = \sum_{k \in \mathcal{K}} \xi_k$
            \State $n \gets n+1$
        \EndWhile 
    \end{algorithmic}
\end{algorithm}

Note that the transmit and relay precoders in \eqref{eq:precoder_optimal_bs} and \eqref{eq:precoder_optimal_re}, respectively, are given in terms of one another. We therefore propose the alternating-optimization strategy in Algorithm~\ref{algorithm}, which minimizes the WMSE by iteratively optimizing the objective function for the receiver structures in \eqref{eq:mmse_receiver}, the MSE weight coefficients in \eqref{eq:wmse_optimal_weight}, and transmit and relay precoders in \eqref{eq:precoder_optimal_bs} and \eqref{eq:precoder_optimal_re}. The algorithm is assumed to converge when the power of the difference for two consecutive precoders is sufficiently small. Note that the value of the objective function increases through iterations due to the power constraints, and the proposed algorithm, hence, eventually converges to a limit value.
Following similar steps as in \cite[Section IV-A]{Cioffi2008_WSR} and  \cite{Kaleva2016_DecenSumRate}, one can prove the convergence in full detail. 

\section{Simulation Results} \label{sec:results}

\begin{table}[!t]
    \vspace{-0.1in}
	\caption{Simulation Parameters} \vspace{-0.1in}
	\label{tab:simulation_parameters}
	\centering
	\begin{tabular}{lc}
		\hline
		Parameter & Value \\
		\hline\hline
        User angular position $([\theta_\mathsf{min},\theta_\mathsf{max}])$ & $[{-}60^\circ,60^\circ]$ \\
        Minimum distance $(d_\mathsf{min})$ & $50\,\text{m}$\\
		Maximum distance $(d_\mathsf{max})$ & $\{150,250\}\,\text{m}$\\
		\# of channel clusters {\color{black}($M_\mathsf{sr}$, $M_\mathsf{rd}$)} & $4$ \\
        \# of channel rays ({\color{black}$L_\mathsf{sr}$, $L_\mathsf{rd}$}) & $5$ \\
        Angular spread of rays & $10^\circ$\\
        Angular spread of clusters & $40^\circ$\\
        Antenna gain at BS \& relay & $8\,\text{dBi}$\\
        Noise figure & $9\,\text{dB}$ \\
        Signal bandwidth & $100\,\text{MHz}$ \\
        Thermal noise & ${-}174\,\text{dBm/Hz}$\\ 
        Error tolerance $(\epsilon)$ & $0.001$ \\
        Frequency $(f_{\rm c})$ & $28\,\text{GHz}$ \\
        \color{black}Maximum number of iterations ($n_\mathsf{max}$) & \color{black} 200 \\
        \color{black} Antenna element spacing ($d$) & \color{black} $\lambda/2$\\
		\hline
	\end{tabular} \vspace{-0.2in}
\end{table}

In this section, we present the numerical results for the proposed sum-rate maximizing precoding strategy along with those of zero forcing (ZF) and regularized ZF (RZF) precoders for comparison purposes. We adopt the 3GPP mmWave channel model for urban micro (UMi) environment with the path loss $\mathsf{PL}(x) \,{=}\, 32.4 \,{+}\, 21\log_{10}(x) \,{+}\, 20\log_{10}(f_{\rm c})$, where $x$ is the LoS distance, and $f_{\rm c}$ is the carrier frequency \cite{3GPP_TR38901}. The complete list of simulation parameters is given in Table~\ref{tab:simulation_parameters}.

\begin{figure}[!t]
\vspace{-0.25in}
\centering
\hspace*{-0.25in}
\includegraphics[width=0.5\textwidth]{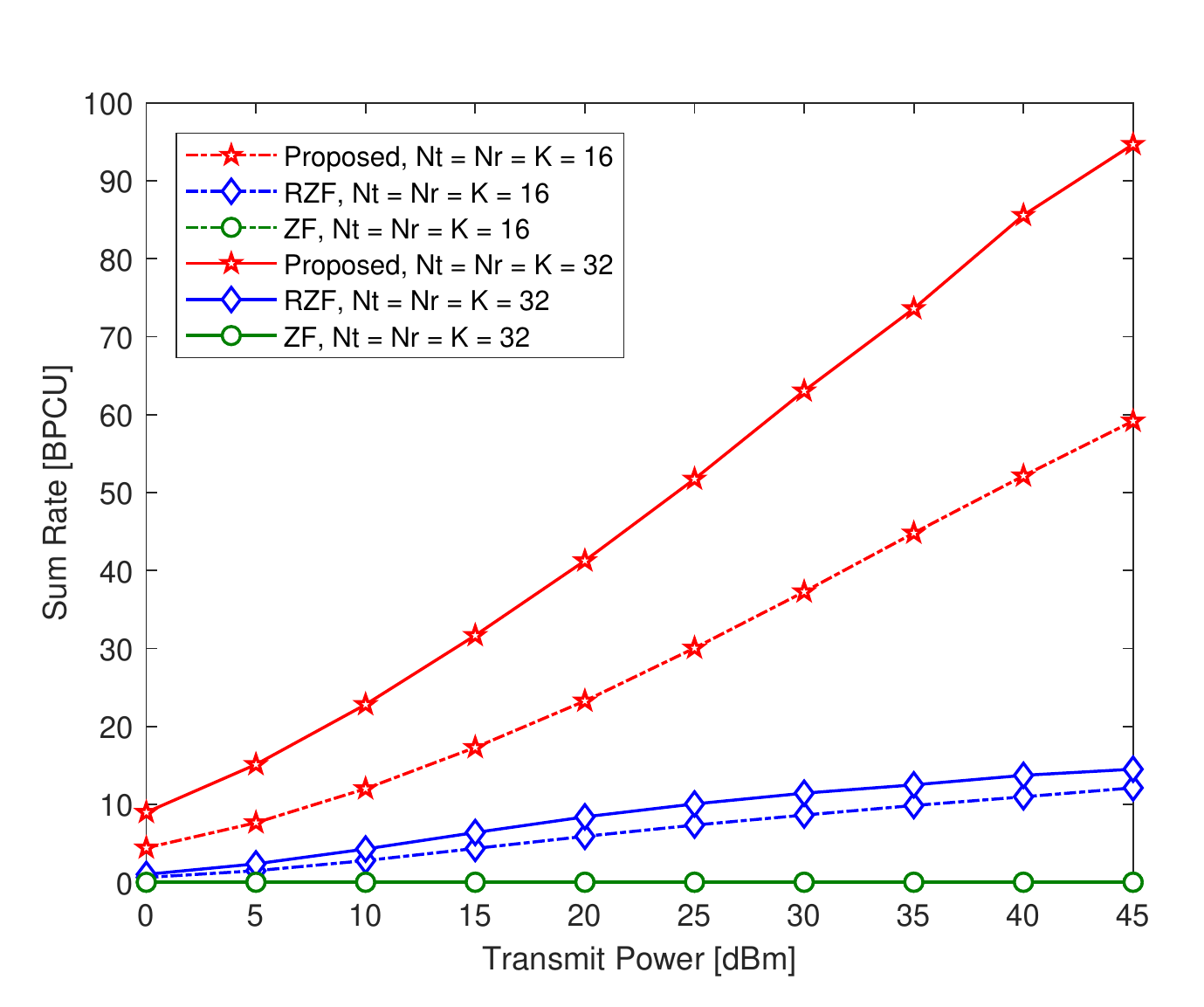}
\vspace{-0.15in}
\caption{Sum rate vs. transmit power for $d_\mathsf{r} \,{=}\, d_\mathsf{min} \,{=}\, 50\,\text{m}$, $d_\mathsf{max} \,{=}\, 150\,\text{m}$, $N_\mathsf{t}\,{=}\,N_\mathsf{r}\,{=}\,K\,{\in}\,\color{black}\{16,32\}$.}
\label{fig:rate_vs_power}
\end{figure}

In Fig.~\ref{fig:rate_vs_power}, we depict the sum rate along with the transmit power with $\mathsf{P}_\mathsf{bs}\,{=}\,\mathsf{P}_\mathsf{re}$. We assume that the relay node is located at the minimum user distance, i.e., $d_\mathsf{r} \,{=}\, d_\mathsf{min} \,{=}\, 50\,\text{m}$, and $d_\mathsf{max} \,{=}\, 150\,\text{m}$ for  $N_\mathsf{t}\,{=}\,N_\mathsf{r}\,{=}\,K\,{\in}\,\color{black}\{16,32\}$. {\color{black}This scenario well represents a densely packed mmWave network which serves as many users as the number of transmit (and relay) antennas, and hence the system can be categorized as \textit{overloaded}}. We observe that the sum-rate performance of the proposed strategy is much better than both RZF and ZF, with the performance gap increasing drastically as the number of antennas and users gets larger. Note that {\color{black}the sum-rate} performance of RZF decreases with increasing number of users (due to larger multiuser interference) especially at high power, while the sum rate of the proposed strategy, in contrast, improves even further. {\color{black}Note also that ZF cannot achieve any positive sum rate under this overloaded scenario.}     
\begin{figure}[!t]
\vspace{-0.3in}
\centering
\hspace*{-0.25in}
\includegraphics[width=0.5\textwidth]{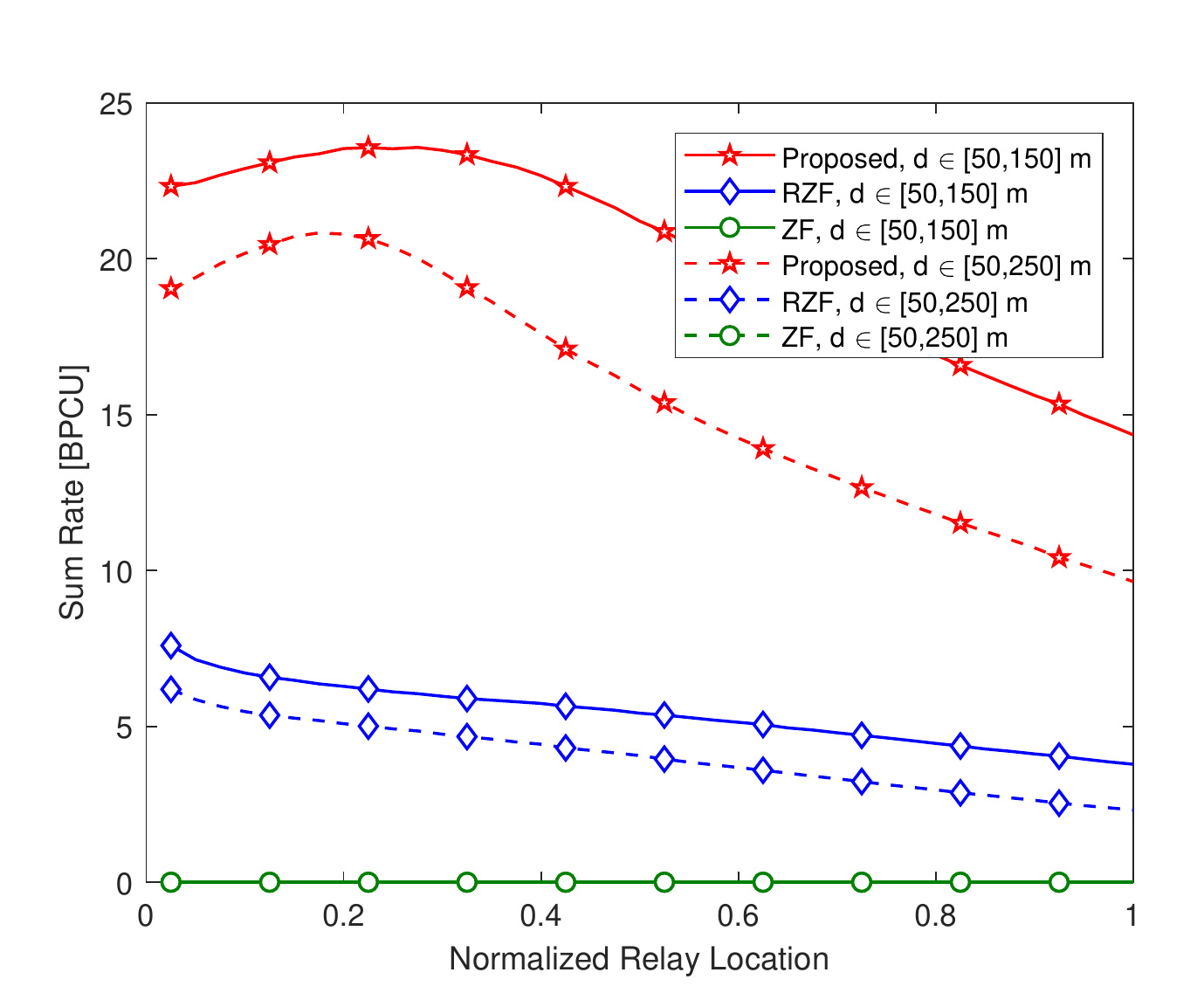}
\vspace{-0.15in}
\caption{Sum rate vs. normalized relay location (i.e., $d_\mathsf{r}/d_\mathsf{max}$) for $d_\mathsf{min} \,{=}\, 50\,\text{m}$, $d_\mathsf{max} \,{=}\, \{150,250\}\,\text{m}$, $N_\mathsf{t}\,{=}\,N_\mathsf{r}\,{=}\,K\,{=}\,{\color{black}16}$, and $\mathsf{P}_\mathsf{bs}\,{=}\,\mathsf{P}_\mathsf{re}\,{=}\,20\,\text{dBm}$. }
\label{fig:rate_vs_location}
\vspace{-0.15in}
\end{figure}

\begin{figure}[!t]
\vspace{-0.25in}
\centering
\hspace*{-0.25in}
\includegraphics[width=0.5\textwidth]{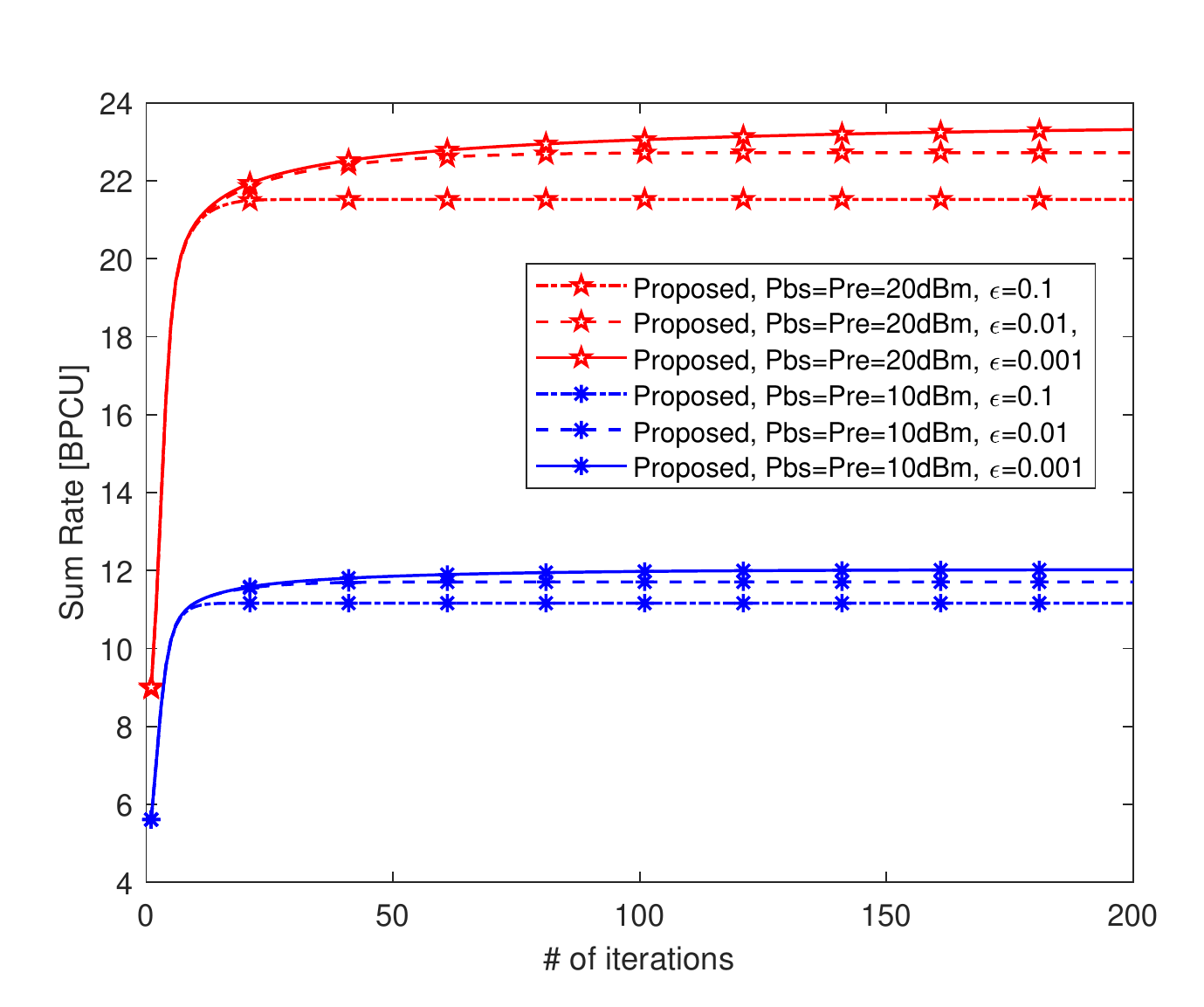}
\vspace{-0.15in}
\caption{{\color{black}Sum rate vs. number of iterations for $d_\mathsf{r} \,{=}\, d_\mathsf{min} \,{=}\, 50\,\text{m}$, $d_\mathsf{max} \,{=}\, 150\,\text{m}$, $N_\mathsf{t}\,{=}\,N_\mathsf{r}\,{=}\,K\,{=}\,16$, $\mathsf{P}_\mathsf{bs}\,{=}\,\mathsf{P}_\mathsf{re}\,{=}\,20\,\text{dBm}$, $\epsilon \,{\in}\, \{0.1,0.01,0.001\}$.}}\vspace{-0.15in}
\label{fig:rate_vs_iteration}
\end{figure}

In Fig.~\ref{fig:rate_vs_location}, we investigate the sum-rate performance for normalized relay location $d_\mathsf{r}/d_\mathsf{max}$ assuming $d_\mathsf{max} \,{=}\, \{150,250\}\,\text{m}$, $\mathsf{P}_\mathsf{bs}\,{=}\,\mathsf{P}_\mathsf{re}\,{=}\,20\,\text{dBm}$, and $N_\mathsf{t}\,{=}\,N_\mathsf{r}\,{=}\,K\,{=}\,{\color{black}16}$. We observe for any $d_\mathsf{max}$ that the best sum rate is obtained for RZF when the relay node appears to be as close to the BS as possible. This result implies that the relay-aided transmission loses its practicality for the RZF precoder as the respective performance maximizes when the relay node is roughly co-located with the BS. The optimal performance for the proposed precoding scheme, in contrast, requires the relay node to be placed off the BS by roughly  $d_\mathsf{r}^{\,\mathsf{opt}} \,{\approx}\, d_\mathsf{min}$. {\color{black}In addition, we present the convergence behavior of the proposed algorithm in Fig.~\ref{fig:rate_vs_iteration} along with various error tolerance values of $\epsilon \,{\in}\, \{0.1,0.01,0.001\}$. We observe that the proposed algorithm converges after as low as $10$ iterations depending on the particular choice of the error tolerance $\epsilon$ and transmit power.}

\section{Conclusion} 
\label{sec:conclusion}

We consider a sum-rate maximizing joint precoder design for a relay-aided multiuser mmWave scenario. Resorting to WMMSE optimization, we obtain closed forms of the precoders and compute them through alternating-optimization iterations. The numerical results verify the superiority of the proposed scheme as compared to RZF and ZF schemes.
\begin{appendices}

\section{Proof of Theorem~\ref{theorem} } \label{appendix}
We first compute the derivative of $\mathcal{L}(\textbf{F},\textbf{G},V_k)$ in \eqref{eq:lagrangian} with respect to $V_k$, and consider the respective roots which yields 
\begin{align}
\sqrt{ \rho_\mathsf{s} \rho_\mathsf{r} } v_k \textbf{f}_k^{\rm H} \textbf{H}_\mathsf{sr} ^{\rm H} \textbf{G}^{\rm H}  \textbf{h}_k^{\rm H} &= \rho_\mathsf{s} \rho_\mathsf{r} \sum_{i \in \mathcal{K}} v_k  \textbf{h}_k \textbf{G} \textbf{H}_\mathsf{sr}  \textbf{f}_i \textbf{f}_i^{\rm H} \textbf{H}_\mathsf{sr} ^{\rm H} \textbf{G}^{\rm H}  \textbf{h}_k^{\rm H} V_k  \nonumber \\
& \hspace{-0.1in} + \rho_\mathsf{r} v_k \sigma^2_\mathsf{r} \textbf{h}_k \textbf{G} \textbf{G}^{\rm H}  \textbf{h}_k^{\rm H} V_k +  \sigma^2_\mathsf{d} v_k V_k , \label{eq:appendix_receiver}
\end{align}
which obtains \eqref{eq:mmse_receiver}. We then take the gradient of \eqref{eq:lagrangian} with respect to ${\textbf{f}_k^{\rm H}}$ and consider the respective roots as follows 
\begin{align}
\sqrt{ \rho_\mathsf{s} \rho_\mathsf{r} } v_k V_k^* \textbf{H}_\mathsf{sr} ^{\rm H} \textbf{G}^{\rm H}  \textbf{h}_k^{\rm H} & =  \rho_\mathsf{s} \rho_\mathsf{r} \sum_{i \in \mathcal{K}}  v_i |V_i|^2 \textbf{H}_\mathsf{sr} ^{\rm H} \textbf{G}^{\rm H}  \textbf{h}_i^{\rm H}  \textbf{h}_i  \textbf{G} \textbf{H}_\mathsf{sr}  \textbf{f}_k \nonumber \\
&\quad + \beta_1\textbf{f}_k  +  \rho_\mathsf{s}  \beta_2  \textbf{H}_\mathsf{sr} ^{\rm H} \textbf{G}^{\rm H} \textbf{G} \textbf{H}_\mathsf{sr}  \textbf{f}_k , \label{Pk_power}
\end{align}
which readily produces \eqref{eq:precoder_optimal_bs}.

We finally take the gradient of \eqref{eq:lagrangian} with respect to $\textbf{G}^{\rm H}$ to obtain \eqref{eq:precoder_optimal_re}, which similarly yields
\begin{align}
\sqrt{ \rho_\mathsf{s} \rho_\mathsf{r}} \sum_{i \in \mathcal{K}} v_i V_i^* \textbf{h}_i^{\rm H} \textbf{f}_i^{\rm H}\textbf{H}_\mathsf{sr} ^{\rm H} & = \left( \rho_\mathsf{r} \sum_{k \in \mathcal{K}} v_k |V_k|^2 \textbf{h}_k^{\rm H}   \textbf{h}_k + \beta_2 \textbf{I}_K \right) \nonumber \\
& \hspace{-0.1in} \times \textbf{G} \left( \rho_\mathsf{s} \textbf{H}_\mathsf{sr}  \textbf{F}  \textbf{F}^{\rm H} \textbf{H}_\mathsf{sr} ^{\rm H}  + \sigma^2_\mathsf{r} \textbf{I}_{N_\mathsf{r}} \right) . \label{G_power}
\end{align} 

In order to derive \eqref{eq:beta2} for $\beta_2$, we first post-multiply both sides of \eqref{eq:appendix_receiver} by $V_k^*$, and compute the summation of the both sides over all user indices. We then pre-multiply \eqref{G_power} by $\textbf{G}^{\rm H}$, compute the trace of both equations, and compare them to readily obtain \eqref{eq:beta2}. Similarly, to calculate $\beta_1$ in \eqref{eq:beta1}, we post-multiply both sides of \eqref{eq:appendix_receiver} by $V_k^*$, and sum the both sides over all user indices. We pre-multiply \eqref{Pk_power} by $\textbf{f}_k^{\rm H}$, and sum over $k \,{\in}\, \mathcal{K}$. After comparing the trace of both, we obtain \eqref{eq:beta1}. \hfill\IEEEQEDhere

\end{appendices}

\bibliographystyle{IEEEtran}
\bibliography{IEEEabrv,references}

\end{document}